\documentstyle[preprint,aps]{revtex}
\begin{document}
\draft
%
%
\title{
Geometric Phase in Quantum Billiards \\
with a Pointlike Scatterer
}
\author{
Taksu Cheon
}
\address{
Department of Physics, Hosei University, Fujimi, 
Chiyoda-ku, Tokyo 102, Japan
}
\author{
Takaomi Shigehara
}
\address{
Computer Centre, University of Tokyo, Yayoi, 
Bunkyo-ku, Tokyo 113, Japan 
}
\date{July 28, 1995}
%
\maketitle
\begin{abstract}
We examine the quantum energy levels of rectangular billiards 
with a pointlike scatterer in one and two dimensions.  
By varying the location and the strength of the scatterer, 
we systematically find diabolical degeneracies among 
various levels.  The associated Berry phase is illustrated, 
and the existence of localized wave functions is pointed out.  
In one dimension, even the ground state is shown to display 
the sign reversal with a mechanism to circumvent 
the Sturm-Liouville theorem.
\end{abstract}
%
%
%
%
\narrowtext
A sense of mystery has accompanied the geometric phase
from the day of its discovery \cite{BE84,AA87}.
It has deepened when the connection between 
the geometric phase and the gauge field has been 
recognized \cite{WZ84}.
Although there are several model systems
in relatively simple settings, few qualify as transparent
examples that give us the intuition into 
the origin and the workings of the geometric phase.

In this Letter, we examine the candidates for 
just such examples. 
We study the bound state problem in a 
rectangular billiard with a pointlike scatterer inside.
The zero-size scatterer is an idealized limit
of a small obstacle. In this limit, the solution 
of the problem is expected to be simplified substantially. 
That simplification is indeed
realized when the divergence caused by the singularity 
of the interaction is handled properly with the self-adjoint 
extension theory of functional analysis
\cite{AG88,ZO80,SZ91,SH94}.   
We look at the energy level structure
of this  system along with its one-dimensional analogue. 
We point out the existence of a set of diabolical points 
and of associated geometric phase.  
An examination of wave functions around the diabolical 
points also reveals the emergence of unexpectedly 
localized wave functions.

	We first consider a one-dimensional delta
potential problem with the Dirichlet 
boundary condition (vanishing wave function) 
at $x = 0$ and $1$.  We place 
a pointlike scatterer of strength $1/s$ at $x = x_0$:
\begin{eqnarray}
\label{1}
V = {1 \over s}\ \delta (x-x_0).
\end{eqnarray}
The diagonal element of the free Green's function 
at $x = x_0$ is given by
\begin{eqnarray}
\label{2}
G(\omega ) = \sum\limits_{n=1}^\infty 
{{{\varphi _n^2(x_0)} \over {\omega -\varepsilon _n}}}
\end{eqnarray}
where $\varepsilon_n=\pi /4\cdot n^2$ and
$\varphi_n (x)=\sqrt{2}\sin(n\pi x)\  (n=1, 2,...)$
are the eigenvalues and 
eigenfunctions of the unperturbed system.  
(We chose the mass to be $2\pi$.)  The 
eigenvalues of the problem with the delta function 
are given as the solutions of the 
equation
\begin{eqnarray}
\label{3}
G(\omega )-s = 0.
\end{eqnarray}
The eigenfunction (apart from the normalization) 
corresponding to an eigenvalue $\omega_\alpha$ is 
given by
\begin{eqnarray}
\label{4}
\psi _\alpha (x) 
= \sum\limits_{n=1}^\infty {{{\varphi _n(x_0)} \over 
{\omega _\alpha -\varepsilon _n}}\varphi _n(x)}.
\end{eqnarray}

	In Fig.1, we plot the first six eigenvalues 
at several $x_0$ as a function of $s$.  Fig. 1 
(a) shows the existence of level crossings at 
infinite strength $s$ = 0 with $x_0$ = 0.5.  Fig. 1 (b) 
displays avoided crossings around $s$ = 0 for $x_0$ = 0.47, 
indicating that the crossings 
at $x_0$ = 0.5 are diabolical.  That this is indeed the case 
is proved by examining Fig. 2 
(a), where we display wave functions at various 
values of $(x_0, s)$ around the point 
(0.5, 0) starting from the ground state at (0.5, 0.1).  
By going round the point (0.5, 0), 
the wave function experiences the Herzberg-Longuet-Higgins 
sign reversal \cite{HL63}, which 
is a special case (co-dimension two) of the Berry's
geometric phase for 
real symmetric Hamiltonians.  The mechanism 
of the sign reversal in this simplest of all examples 
is easy to comprehend:  It is a 
consequence of two conflicting ways of connecting 
nodeless and one-node wave functions.  

At first sight, this result seems to 
contradict with the Sturm-Liouville theorem
of no-crossing and no-node \cite{CH31}, 
which states that in one dimension 
there is no level crossing and the 
ground state always stays nodeless.  However, at the 
diabolical point $(x_0, s) = (0.5, 0)$, 
the strength of the potential is infinite, where 
the no-crossing theorem does not have 
to hold.  Also, as seen in Fig. 1, the ground state 
at $s>0$ side is smoothly connected to 
the second state with one node at $s < 0$ side.  
This happens because when one goes 
from positive to negative $s$, a disconnected 
level appears from the negative infinity to become 
the ground state for $s < 0$.  In this rather 
tricky manner, the ground state of $s > 0$ side is 
permitted to acquire the sign reversal through the 
rotation around the diabolical point.

	Looking at Fig. 1 (a), we notice that 
there are a series of pairwise degeneracies at 
$(x_0, s) = (0.5, 0)$.  This occurs between 
the states with even $n$ which are unaffected by 
the delta potential at the center and the states 
with odd $n$ which lie next to them.   
Similarly, one realizes from Fig. 1 (c)-(e)  that 
the third state, sixth state, ninth state,  {\em etc.} 
get crossed by the states
next to them  at $(x_0, s) = (1/3, 0)$.  More generally, 
if one assumes that $N$ and $M$ are the relative primes 
satisfying $N > M > 0$, the location 
$(x_0, s) = (M/N, 0)$ is a diabolical point for 
the $(N-1)$-th and $N$-th states, $(2N-1)$-th 
and $2N$-th states, {\em etc.} (all counted at $s > 0$ side).  
Clearly, one exhausts all the level 
crossings of the system in this manner.  
We show in Fig. 2 (b), the morphosis of the 
eighth wave function around the diabolical point 
$(x_0, s) = (1/9, 0)$.  A noteworthy 
feature is in the middle right figure 
($s_0\simeq 0$ and $x_0$ slightly larger than $1/9$), 
where one observes a wave function highly localized 
between one edge and the location of the 
delta potential.  One can generalize this result:  
By placing a strong delta potential at an 
appropriate location, one can obtain a wave 
function which is localized arbitrary sharply 
near the edge.

We next treat the case of two dimension.
We consider a quantum particle of mass 
$2\pi$ moving inside a rectangle surrounded by the 
boundaries $x = 0$, $x = R$, $y = 0$ and 
$y = 1/R$, on which the wave function is assumed 
to vanish.
The eigenstates for this empty billiard
problem is given by
$\varphi_{nm}(x,y)=2\sin (n\pi x/R)\cdot \sin (m\pi y R)$
and their energies
$\varepsilon_{nm} = \pi /4 \cdot (n^2/R^2+m^2R^2) 
\ ( n,m=1,2...)$. 
We place a pointlike scatterer at $(x_0, y_0)$.  
Naively, we would have the two-dimensional analogue of
eqs. (\ref{1}) - (\ref{3}).
However, this is not possible since 
the infinite sum
\begin{eqnarray}
\label{5}
G(\omega ) = \sum\limits_{n,m=1}^\infty 
{{{\varphi _{nm}^2(x_0,y_0)} \over {\omega -\varepsilon _{nm}}}}
\end{eqnarray}
diverges logarithmically because of the constant density
of states per unit energy.
Instead, we resort to the self-adjoint extension theory of 
functional analysis.  We refer to Ref. \cite{SH94} 
for a fuller treatment of the problem, and 
simply start from the results obtained there.  
The eigenvalues of the full 
problem are given by the solutions of
\begin{eqnarray}
\label{6}
\overline G(\omega )-\bar s = 0
\end{eqnarray}
where  
\begin{eqnarray}
\label{7}
\overline G(\omega ) = \sum\limits_{n,m=1}^\infty 
{\varphi _{nm}^2(x_0,y_0) 
\left[ {{1 \over {\omega -\varepsilon _{nm}}}
+{{\varepsilon _{nm}} \over {\varepsilon _{nm}^2+1}}} \right]}
\end{eqnarray}
is the regularized version of $G(\omega )$, and  
$\bar s$, which we call {\em formal} inverse strength, 
is a real number  specifying the nature of the  pointlike
scatterer.  The equation (\ref{6}) can be interpreted as 
the renormalized eigenvalue equation if $\bar s$ is identified 
as the inverse of the renormalized coupling strength.
In one dimension, we can consider an analogous 
equations to eqs.(\ref{6}) and(\ref{7}). 
In this case, however, each term in the bracket of eq.(\ref{7}) 
converges separately.  This allows us 
to define the inverse strength $s$ in terms of $\bar s$  
as a finite quantity.
This is the reason why we are able to deal with 
the delta potential in one dimension without any difficulty. 
The eigenfunction corresponding to a 
solution $\omega_\alpha$ of eq.(\ref{6}) is given by
\begin{eqnarray}
\label{8}
\psi _\alpha (x,y) = \sum\limits_{n,m=1}^\infty 
{{{\varphi _{nm}(x_0,y_0)} \over 
{\omega _\alpha -\varepsilon _{nm}}}\varphi _{nm}(x,y)}.
\end{eqnarray}

We look at the energy of the system as 
a function of the location of the 
scatterer, $(x_0, y_0)$ with fixed value of $\bar s$.  
Because of the mirror symmetry of the 
system, it is sufficient to consider the area 
$x_0\in  [0, R/2]$ and $y_0\in  [0, 1/(2R)]$.  In 
Fig. 3, we show the energy of the lowest eleven 
levels as a function of $y_0$ with $x_0 = 0.5R$
for Fig. 3 (a), and with $x_0 = 0.48R$ for  Fig. 3 (b).  
The value of $R$ is set to 
$\pi /e \simeq 1.1557273$.  The formal inverse strength is 
chosen to  $\bar s = 1/10$.  Six diabolical 
crossings can be spotted in Fig. 3 
(including the one between the eighth and ninth states 
which might require further magnification to see 
the avoided crossing off $x_0 = 0.5R$).  
All the crossings involve a state with even $n$, 
which is unaffected by the presence of 
the scatterer because of its node line along $x_0 = 0.5R$.  
As a result, these states appear 
flat in Fig. 3 (let us call these flat energies of 
unaffected states $\varepsilon_\varphi$ collectively).  

In general, the level crossing in the system
with pointlike scatterer can occur 
only when one of the eigenstates is 
unaffected by the scatterer.  This occurs 
either along the line $x_0 = (M/N)R$ or 
along $y_0 = M/(NR)$ where $N$ and $M$ are the relative primes 
satisfying $N > M >0$.  
There, unperturbed wave functions with $(N-1)$ nodes 
in $x$ (or $y$) direction become the 
solutions of the full problem.  The other coordinate 
$y_0$ (or $x_0$) of the crossing point is 
then determined by eq. (\ref{6}) with 
$\omega = \varepsilon_\varphi$.  Thus we have 
a systematic way of locating all 
diabolical crossings.  As we vary $R$, the diabolical 
locations move along $x_0 = (M/N)R$ or $y_0 = M/(NR)$.  
An interesting case is the right square $R = 1$, where the 
lowest diabolical point (between the second and 
third states) moves to the center $(x_0, y_0)$ = $(0.5, 0.5)$.  
We observe {\em no sign reversal} around this point because 
this and the other diabolical point at its mirror 
location $y'_0$ = $1/R-y_0$ merge to form a degenerate 
diabolical point with the Berry phase $2\pi$.

	Going back to the specific case of $R = \pi /e$,
we display, in Fig.4, the profile of eigenfunctions 
around two of the diabolical points; 
(a) $(x_0, y_0)$ = $(0.5R, 0.3177/R)$ involving the second 
and third states and 
(b) $(x_0, y_0)$ = $(0.5R, 0.3875/R)$ for the tenth and eleventh 
states.  In case of (a), the third state keeps the 
unperturbed quantum numbers $(n, m) = (2, 1)$ 
and the energy $\varepsilon_\varphi \simeq  3.4$, while in 
case of (b), the tenth state has $(n, m) = (4, 2)$ 
and $\varepsilon_\varphi \simeq 13.6$.  As in one dimension, 
we learn from these examples that the sign reversal 
is a result of two conflicting ways 
to connect wave functions with different node structures. 
One can also look at this sign reversal as a rotation
of the wave function by 180 degree while the pointlike
scatterer makes the full turn of 360 degree.  
As in one dimension, 
we observe the appearance of localized wave functions 
around the diabolical level 
crossings.  Its possible relevance to the ``scaring'' 
phenomena \cite{HE84,BO89} is an interesting 
open problem.

        Finally, we view our findings from a perspective 
and discuss their potential utilities.  
It is important to notice that the rectangular 
shape of the billiard is a matter of our 
technical choice than the necessary condition for 
the occurrence of the diabolical 
crossings.  In fact, by linearizing 
$\overline G(\omega)$ in the
neighborhood of $\varepsilon_\varphi$  in eq. (\ref{6}), 
we obtain a second-order algebraic equation, with 
which we can prove the appearance of the Berry phase 
around diabolical points without any reference to 
unperturbed eigenfunctions.  Therefore, we conclude 
that a generic system with a pointlike scatterer 
possesses diabolical crossings and the associated 
Berry phase.  Also, the diabolical crossing discussed 
here is not a result of the mathematical abstraction 
inherent in a pointlike scatterer.  
Indeed, we can show through  
numerical Fourier-basis diagonalization that the 
sign reversal is observed in a billiard with a polygonal 
obstacle of finite size such as the one
discussed in Ref. \cite{CC89}. When the size of the
obstacle is small, the diabolical crossing occurs at a
location close to the one predicted for a pointlike
scatterer. 
In this case, however, there is no systematic way to
locate the diabolical points exactly.
It is the very pointlike nature of the obstacle 
that allows the simple expressions such as 
eqs. (\ref{6}) and (\ref{7}). 
We can further hope that the simplicity of the 
formulation of the pointlike scatterer opens
up the possibility of various analytical results
concerning the nature of the diabolical points such 
as their geometrical and statistical properties.

	The primary utility of our examples might 
be a pedagogical one.  This is evident when we compare 
them with the earlier examples of the Berry phase in triangular 
billiards \cite{BW84}.  
The rotation in parameter space in our examples 
literally is the rotation of a pointlike scatterer in the 
coordinate space.  Also, the occurrence of the diabolical 
crossing at the lowest possible states including the ground 
state makes the resulting Berry phase understood in a pictorial, 
intuitive manner.  We can also contemplate a possible use of 
diabolical crossings in mesoscopic devices.  
There may be a situation where an obstacle (or impurity) needs 
to be placed in a certain precise location of the 
device.  This can be achieved by monitoring the degenerate 
resonance energies while moving the obstacle.  
Another more fanciful use may be realized when the phase of 
wave functions becomes utilized.  
A quantum state with required phase can be obtained 
by ``Berry rotating'' an obstacle around its diabolical location.

	We acknowledge our gratitude to Prof. M. V. Berry 
for his suggestion which has led to the conception
of this work. We also thank Prof. H. Terazawa for his
careful reading of the manuscript and useful comments.
T.C. expresses the gratitude  
to the members of the Theory Division of the Institute  
for Nuclear Study, University of Tokyo for offering him
a friendly and stimulating research environment.  
T. S. acknowledges the support of  the Grant-in-Aid for
Encouragement of Young Scientists (No. 07740316) by 
the Ministry of Education,
Science, Sports and Culture of Japan.

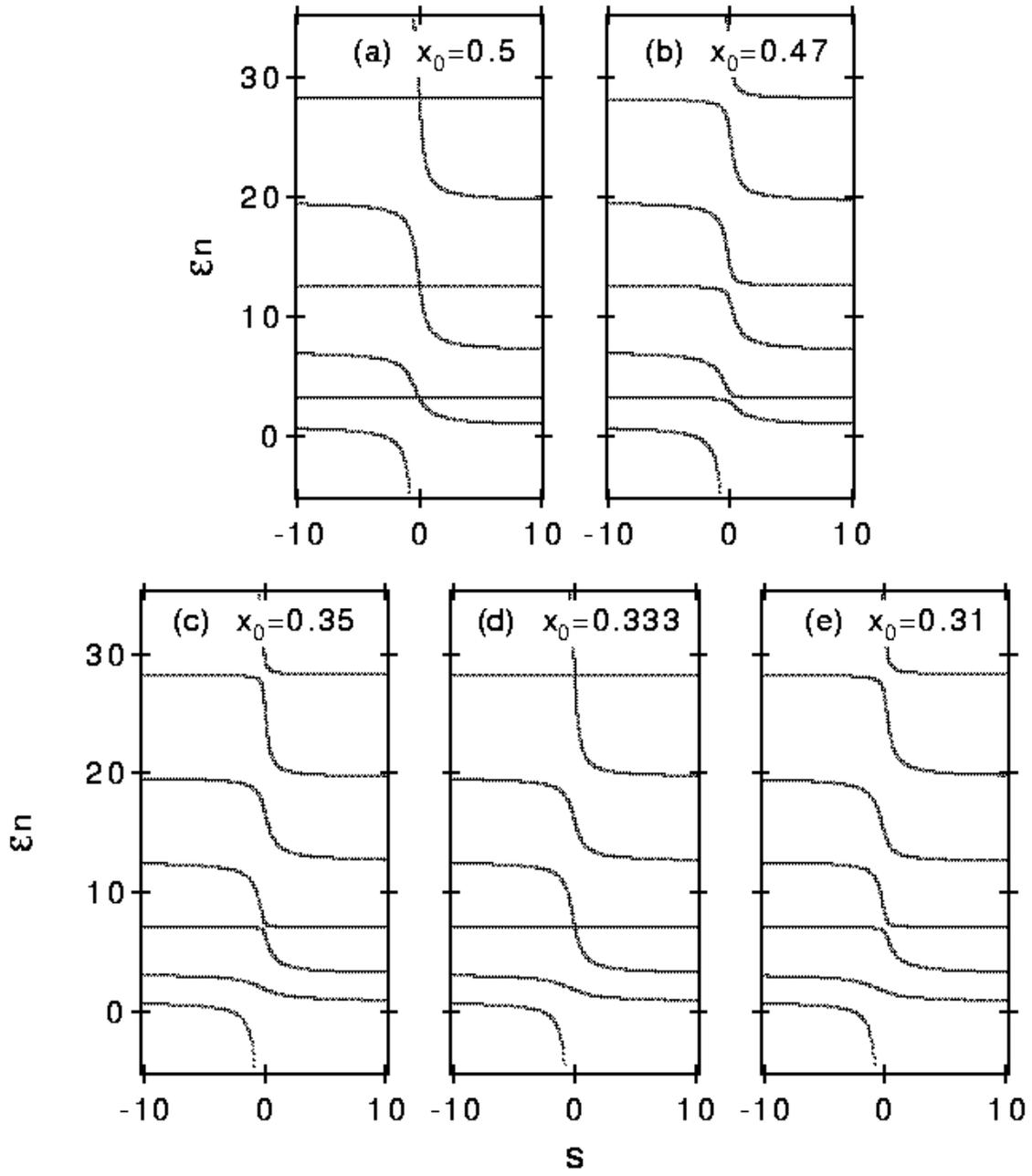
\begin{figure}
\caption{
Six lowest eigenvalues as function of the 
inverse strength of the delta potential 
in one dimension: 
(a) $x_0 = 0.5$, (b) $x_0 = 0.47$, (c) $x_0 = 0.35$,
(d) $x_0 = 0.333$ and (e) $x_0 = 0.31$.
}
\end{figure}
\begin{figure}
\caption{
The change of the wave function eq. (4) 
displaying the Berry phase with 
various values of $(x_0, s)$ around the diabolic points 
(a) $(x_0, s) = (1/2,0)$ and (b) $(x_0, s) = (1/9,0)$ 
}
\end{figure}
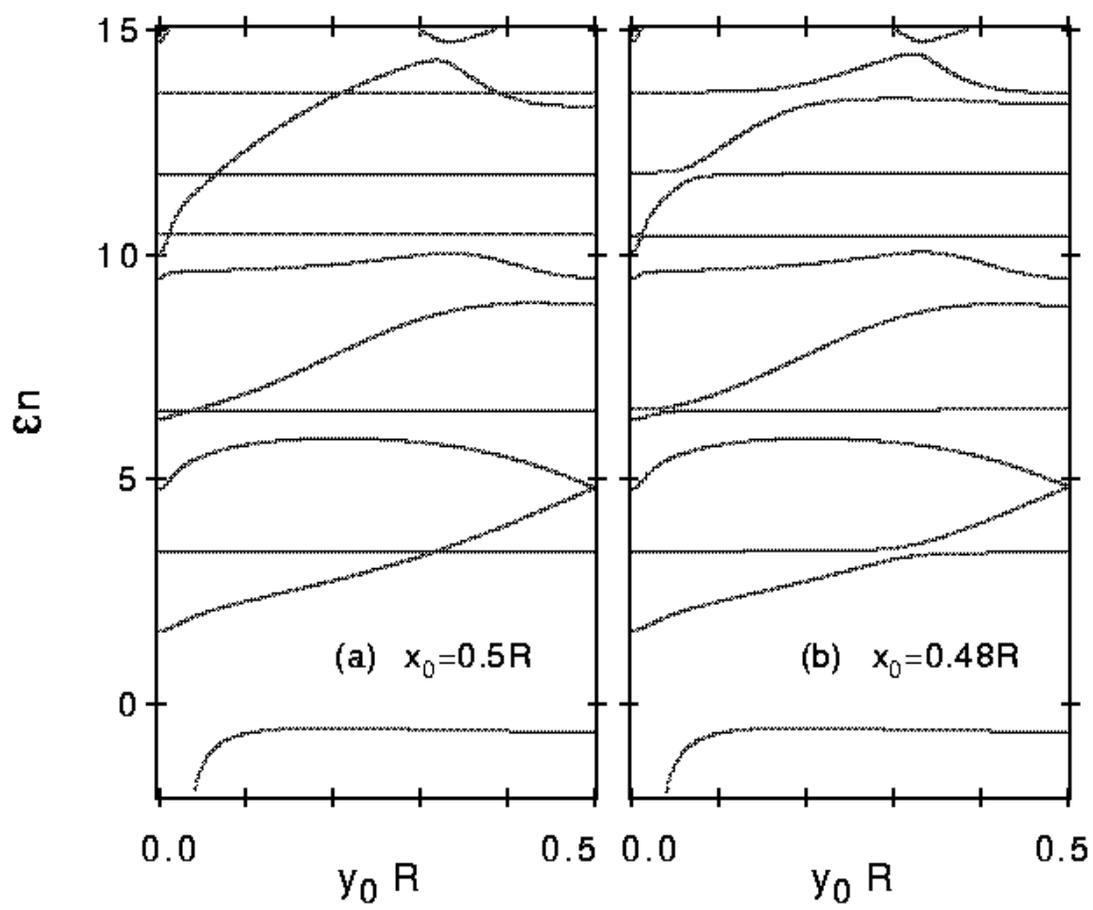
\begin{figure}
\caption{
Eleven lowest eigenvalues as function of the 
perpendicular position of the 
delta potential in two-dimensional billiard while 
keeping its horizontal position 
(a) at the center $x_0 = 0.5R$, and 
(b) slightly off the center $x_0 = 0.48R$.  
Other parameters are set to 
$R = 1.1557273$ and $\bar s = 0.1$.
}
\end{figure}
\begin{figure}
\caption{
The change of the wave function displaying 
the Berry phase around the diabolical points 
(a) $(x_0, y_0)$ = $(0.5R, 0.3177/R)$ and 
(b) $(x_0, y_0)$ = $(0.5R, 0.3875/R)$ 
with $R = 1.1557273$ and $\bar s = 0.1$.  
The solid and broken lines indicate positive and 
negative equilateral lines respectively. 
}
\end{figure}

\begin{thebibliography}{10}

\bibitem{BE84}
M. V. Berry, Proc. Roy. Soc. Lond. {\bf A392}, 43 (1984).

\bibitem{AA87}
Y. Aharonov and J. Anandan, 
Phys. Rev. Lett. {\bf 58}, 1593 (1987).

\bibitem{WZ84}
F. Wilczek and A Zee, Phys. Rev. Lett. {\bf 52}, 2111 (1984).

\bibitem{AG88}
S. Albeverio, F. Gesztesy, R. H{\o}egh-Krohn and H. Holden, 
{\em Solvable Models in Quantum Mechanics}
(Springer, Berlin, 1988).

\bibitem{ZO80}
J. Zorbas, J. Math. Phys. {\bf 21}, 840 (1980).

\bibitem{SZ91}
P. {\v S}eba and K. {\. Z}yczkowski, 
Phys. Rev. {\bf A44}, 3457 (1991).

\bibitem{SH94}
T. Shigehara, Phys. Rev. {\bf E50}, 4357 (1994).

\bibitem{HL63}
G. Herzberg and H. C. Longuet-Higgins, 
Discuss. Faraday Soc. {\bf 35}, 77 (1963).

\bibitem{CH31}
R. Courant and D. Hilbert, 
{\em Methoden der Mathematischen Physik} 
(Springer, Berlin, 1931).

\bibitem{HE84}
E. Heller, Phys. Rev. Lett. {\bf 53}, 1515 (1984).

\bibitem{BO89}
E. B. Bogomolny, Physica {\bf 31D}, 169 (1989).

\bibitem{CC89}
T. Cheon and T. D. Cohen, 
Phys. Rev. Lett. {\bf 62}, 2769 (1989).

\bibitem{BW84}
M. V. Berry and M. Wilkinson, 
Proc. Roy. Soc. Lond. {\bf A392}, 15 (1984).

\end{thebibliography}
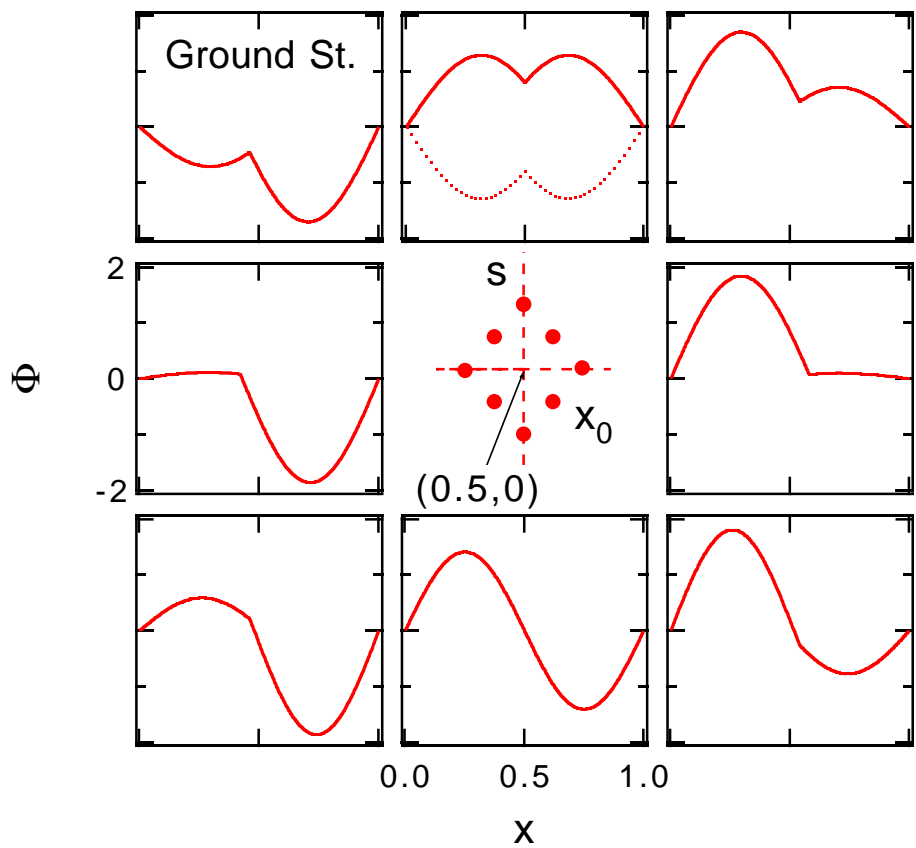
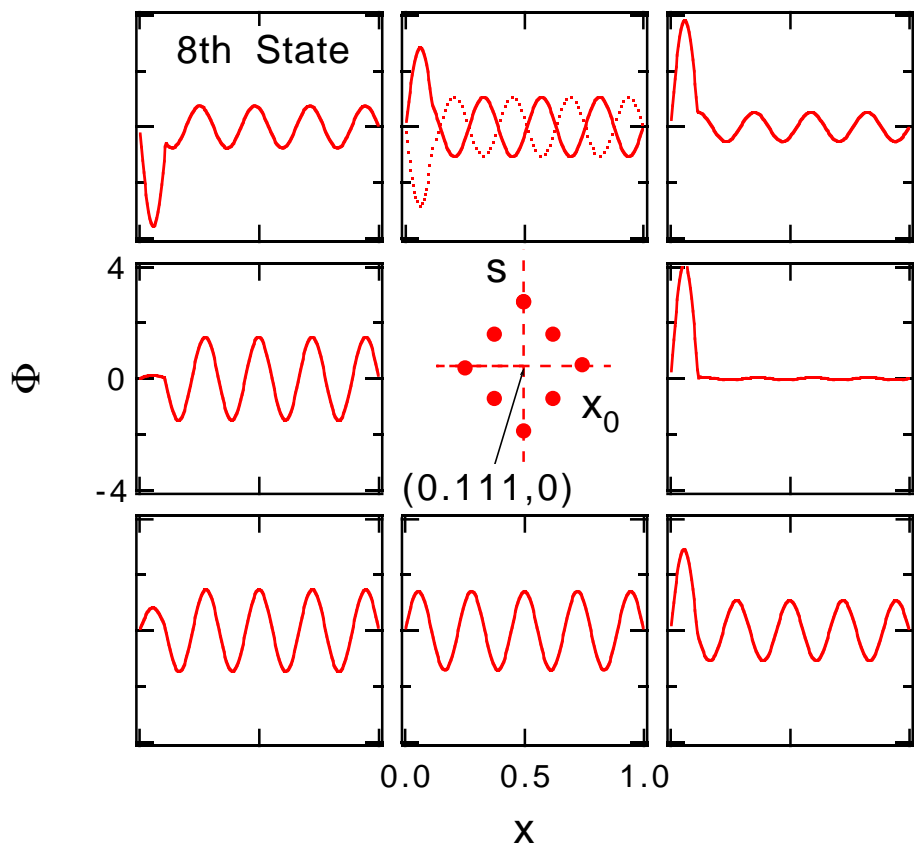
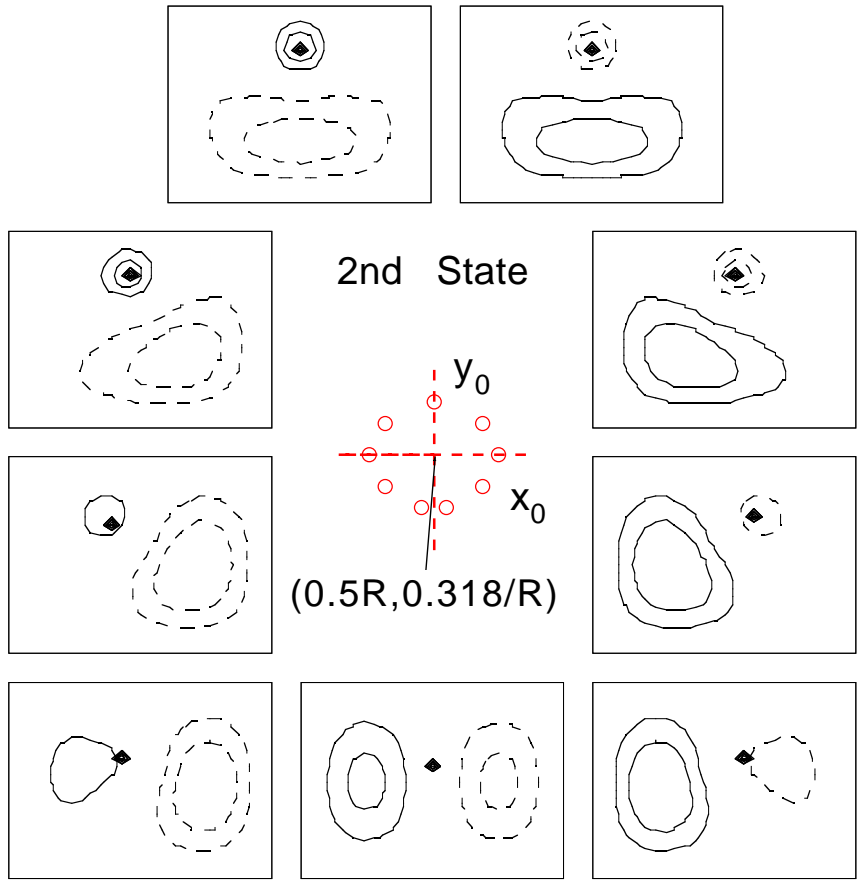
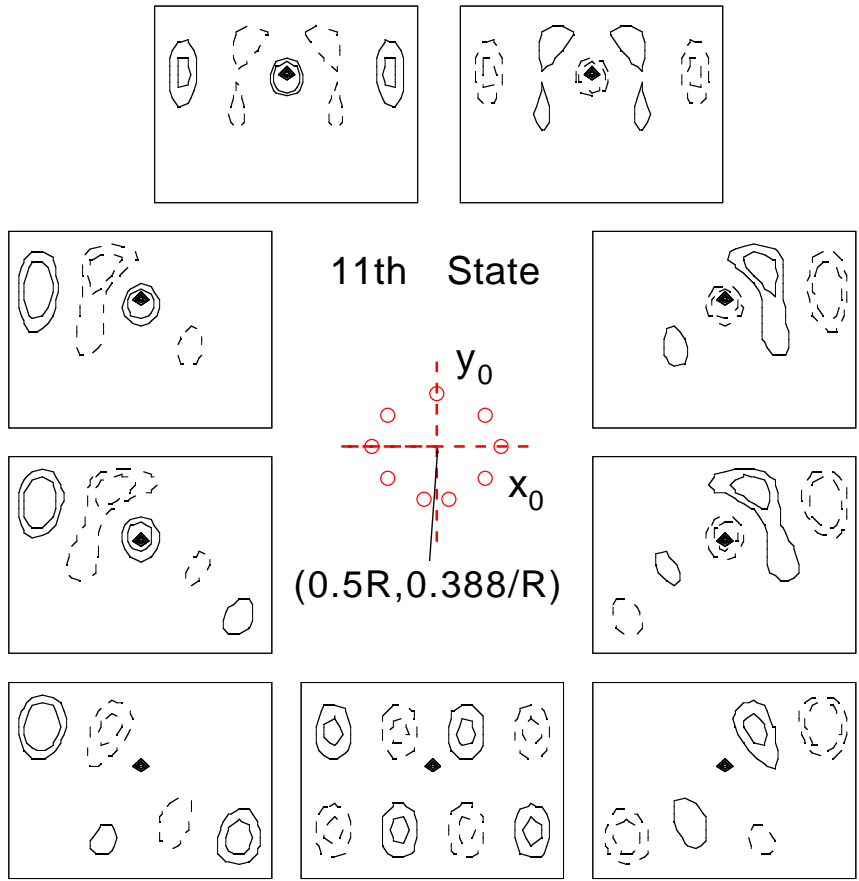
\end{document}